\documentclass[aps,prb, preprint,showpacs,superscriptaddress]{revtex4}
\usepackage{graphicx}
\usepackage{epstopdf}
\usepackage{amsmath,amssymb}
\usepackage{float}

\newcommand{\ignore}[1]{}
\newcommand{\beq}{\begin{equation}}
\newcommand{\eeq}{\end{equation}}
\newcommand{\mbold}[1]{\mbox{\boldmath $ #1 $}}

\begin{document}

\title
{Nontrivial phase diagram for an elastic interaction model of spin crossover materials with antiferromagnetic-like short-range interactions}

\author{Masamichi Nishino}
\email[Corresponding author: ]{nishino.masamichi@nims.go.jp}
\affiliation{International Center for Materials Nanoarchitectonics, National Institute for Materials Science, Tsukuba, Ibaraki 305-0044, Japan} 

\author{Seiji Miyashita} 
\affiliation{Department of Physics, Graduate School of Science, The University of Tokyo, Bunkyo-Ku, Tokyo 113-0033, Japan}

\author{Per Arne Rikvold}
\affiliation{Department of Physics, Florida State University, 
Tallahassee, FL 32306-4350,USA}

\date{\today}

\begin{abstract}

We study the phase diagram of an elastic interaction model for spin crossover (SC) materials with antiferromagnetic-like short-range interactions. 
In this model, the interplay between the short-range interaction and the long-range interaction of elastic origin causes complex phase transitions. 
For relatively weak elastic interactions, the phase diagram is characterized by tricritical points, at which antiferromagnetic (AF) -like  and ferromagnetic (F) -like spinodal lines and a critical line merge. 
On the other hand, for relatively strong elastic interactions, unusual ``horn structures," which are surrounded by the F-like spinodal lines, disorder (D) spinodal lines, and the critical line, are realized at higher temperatures. 
These structures are similar to those obtained in our previous study [Phys. Rev. B {\bf 93}, 064109 (2016)] of an Ising antiferromagnet with infinite-range ferromagnetic interactions, and we find universal features caused by the interplay between the competing short-range and long-range interactions. 
The long-range interaction of elastic origin is irrelevant (inessential) for the critical line. In contrast, the AF-like, F-like, and D spinodal lines result from the long-range interaction of elastic origin. 
This difference causes qualitatively different features of domain formation or nucleation of the new phase: clustering occurs in the former case, while clustering is absent in the latter. Based on the phase diagrams, we discuss the patterns and clustering features of two-step SC transitions, in which the AF-like phase is realized in the intermediate temperature region.

\end{abstract}


\maketitle

----------------------------------------------------------------------------

\section{Introduction}

Spin crossover (SC) compounds exhibit a variety of phase transitions induced by temperature and/or pressure variation, light irradiation, etc.~\cite{Gutlich_book,SC_book2,Gutlich,Kahn,Hauser,Ksenofontov,Bousseksou,Letard,Pillet1,Tanasa,Kimura,Pillet2,Ichiyanagi,Lorenc,Watanabe,Brefuel,Chong,Bousseksou2,Slimani,Collet,Varret,Slimani2,Bertoni,Collet2}. SC transitions have been frequently modeled by Ising-like models with a temperature-dependent field which reflects the different degeneracies of the low spin (LS) and high spin (HS) states. On the atomistic scale, the molecular volume (or size) varies between the LS and HS states through the vibronic coupling between the electronic state and the structure, and thus volume expansion or contraction 
accompanies the phase change of SC crystals.   
  Based on explicit modeling of the molecular size, it was shown that the elastic interaction, which originates from the lattice distortion due to the different molecular sizes,  plays the role of a cooperative interaction and causes SC transitions~\cite{Nishino_elastic,Miya,Nicolazzi1,Enachescu1,Slimani3}. 
The cooperative nature of the elastic interaction has been intensively studied~\cite{Nishino3,Enachescu2,Nicolazzi2,Enachescu3,Kamel_elastic}. 

 The interplay between the short-range and long-range interactions causes complex ordering processes~\cite{Low,Muratov,Sciortino,Giuliani,Nakada1,Nakada2}, and 
it is important for two-step SC transitions~\cite{Koppen,Petrouleas,Jakobi,Real,Boinnard,Buron,Bousseksou3,Nishino_two-step,Kamel,Nishino3,Shatruk,Paez-Espejo}, in which the antiferromagnetic (AF) -like phase mostly appears as a medium-temperature phase. 

It has been shown that the effects of the short-range interaction are essentially different for ferromagnetic (F) -like and AF-like phase transitions at critical points at zero field ($H=0$). 
Namely, the long-range interaction of elasticity is relevant in the ferromagnetic-like transition. It causes the transition to belong to the mean-field universality class~\cite{Miya,Nakada2,Nishino3}, and it works cooperatively with the short-range interaction. 
In contrast, it is irrelevant in the antiferromagnetic-like transition, conserving the Ising universality of the pure short-range model~\cite{Nishino3}. 

It has been shown that infinite-range (Husimi-Temperley) interactions 
can be a good approximation to the long-range nature of the elastic interaction 
in a triangular SC model with short-range frustrated interactions, in which 
a second-order transition with a new critical universality and a BKT transition~\cite{Berezinskii,Kosterlitz} as the end points of the BKT phase have been found~\cite{Nishino_triangular}.  

We have recently studied the square-lattice Ising antiferromagnet with Husimi-Temperley (HT) type long-range ferromagnetic interactions~\cite{Per1} as a simplified model of the square-lattice elastic interaction model with AF-like short-range interactions. 
The Hamiltonian is given by 
\beq
{\mathcal H}= -J \sum_{\langle i,j \rangle} \sigma_i \sigma_j - N \left( H + \frac{A}{2} m \right) m
\label{eq:ham}
\eeq
with $J < 0$ and the long-range interaction strength $A > 0$. 
The variable $\sigma_i$ denotes the spin state of the $i$th site ($-1$ for LS and $+1$ for HS), and $N$ is the number of sites. 
Here, $H$ is an effective field (see Eq.~\eqref{H_eff}). 
It was found that the structures of the phase diagrams in the coordinates of temperature and field, ($T$,$H$), obtained by the mean-field and Monte Carlo (MC) methods for weak and strong long-range interactions are similar, while
for intermediate-strength long-range interactions, the MC simulations show that in the region of $H \ne 0$, tricritical points decompose into pairs of critical end points and mean-field critical points surrounded by horn-shaped regions of metastability. 
The appearance of these unusual ``horn structures" is qualitatively different from the mean-field result.  

An open question remains of whether this kind of structures is realized in the square-lattice elastic interaction model. 
In the present paper, to answer this question, we investigate in detail the phase diagram of the square-lattice elastic interaction model for SC with AF-like short-range interactions. 
First we obtain the phase diagram for relatively weak elastic interactions, and we discuss the properties of the critical points and spinodal lines. 
Next, by considering the shape of the phase diagram, the patterns of typical two-step SC transitions are studied. 
Finally, we investigate the phase diagram for relatively strong elastic interactions. We find unusual ``horn structures," similar to those observed in the Ising antiferromagnet with HT long-range ferromagnetic interactions~\cite{Per1}. 
We study the details of the structures and their origin.

The organization of the rest of the paper is as follows. 
In Sec.~\ref{sec_model}, the model and method are explained. 
In Sec.~\ref{sec_phase_diag_weak}, we show the phase diagram for the SC model with relatively weak elastic interactions, and discuss the types of two-step 
SC transitions. 
In Sec.~\ref{sec_phase_diag_strong}, we present the phase diagram for the SC model with relatively strong elastic interactions, and discuss the origin and properties of the unusual structures that are observed. 
In Sec.~\ref{summary}, we give a summary and suggest topics for future research. 

\section{Model and method}
\label{sec_model}

We study an elastic SC model with antiferromagnetic-like short-range interactions on a square lattice~\cite{Nishino3}, 
in which the lattice can be distorted due to the changes of the positions $\{\mbold{r}_i\}$ and radii $\{R_i\}$ of the molecules. 
The Hamiltonian consists of three terms: elastic interaction (${\cal H}_{\rm Elastic}$), Ising interaction (${\cal H}_{\rm IS}$) and effective field term (${\cal H}_{\rm eff}$), 
\begin{equation}
{\cal H}_{\rm tot}={\cal H}_{\rm Elastic}+{\cal H}_{\rm IS}+{\cal H}_{\rm eff}. 
\label{H_tot}
\end{equation}
Each SC molecule can be in the LS state ($\sigma_i=-1$) or the HS state ($\sigma_i=1$). Its radius $R_i$ is a function of the state $\sigma_i$ because the LS molecule is smaller than the HS molecule, i.e., 
$R_{\rm L}<R_{\rm H}$ where $R_{\rm L}$ ($R_{\rm H}$) is the radius of the LS (HS) molecule. 
The elastic interaction between the nearest-neighbor molecules is given by 
\begin{equation}
{\cal H}_{\rm nn}={k_1 \over 2}\sum _{\langle i,j \rangle} \big[r_{i,j}-(R_i(\sigma_i)+R_j(\sigma_j))\big]^2. 
\label{eq_Hnn}
\end{equation} 
To maintain the square lattice (coordination number), a
small contribution of the following next-nearest-neighbor interaction is necessary: 
\begin{equation}
{\cal H}_{\rm nnn}={k_2 \over 2}\sum _{\langle\langle i,k \rangle\rangle} \big[r_{i,k}-\sqrt{2}(R_i(\sigma_i)+R_k(\sigma_k))\big]^2, 
\label{eq_Hnnn}
\end{equation}
where $k_2=k_1/10$ is set.  
The total elastic interaction is 
\begin{equation}
{\cal H}_{\rm Elastic}={\cal H}_{\rm nn}+{\cal H}_{\rm nnn}. 
\label{ela}
\end{equation}

The nearest-neighbor AF-like Ising interaction (we emphasize that it is in general not of magnetic origin) is considered as the short-range interaction, 
\begin{equation}
{\cal H}_{\rm IS}=-J\sum_{\langle i,j \rangle}\sigma_i \sigma_j, 
\label{H_J}
\end{equation}
where $J<0$.   

In SC systems there is an energy difference ($D=E_{\rm HS}-E_{\rm LS}$) between the HS and LS states. The entropy effect due to the difference of the density of states ($\rho$), whose ratio is defined by $g=\rho_{\rm HS}/\rho_{\rm LS}$, is also important. They are described by an effective field term as 
\begin{eqnarray}
{\cal H}_{\rm eff}=-H \sum_i \sigma_i, 
\label{H_eff}
\end{eqnarray}
where 
\begin{eqnarray}
H \equiv -\frac{1}{2}\left(D-{k_{\rm B}T}\ln g\right). 
\end{eqnarray}

Here we apply a MC method in the $NPT$ ensemble, where the pressure is set to $P=0$. We apply periodic boundary conditions. 
In the MC method, we choose a molecule at the site $i$, and update the spin state 
$\sigma_i$ and the position of the molecule ($x_i,y_i$). 
Then we update the volume of the total system under the condition $P=0$. 
One MC step (MCS) is defined as $L \times L$ repetitions of these procedures, where $L$ denotes the linear dimension (square root of the number of sites) of the system. We apply $10^5$ MCS $-$ $4\times 10^6$ MCS for $L=10-100$ for measurement of the physical quantities in the equilibrium or steady state after the same number of MCS for equilibration.

We study the phase diagram of the model with particular focus on the differences between the 
relatively weak and relatively strong elastic interaction cases, relative to 
the short-range interaction ($J$). 
We set the parameter values as $R_{\rm H}=1.1$, $R_{\rm L}=1$ and $J=-1.0$ (AF-like interaction). 
Here $|J|=1$ is taken as the unit of energy.  
We adopt $k_1=400$ and $k_2=40$ for the former case and $k_1=1600$ and $k_2=160$ for the latter case, in both of which the ground state in zero field is the AF-like phase~\cite{Nishino3}.

\begin{figure}
\centerline{\includegraphics[clip,width=10.0cm]{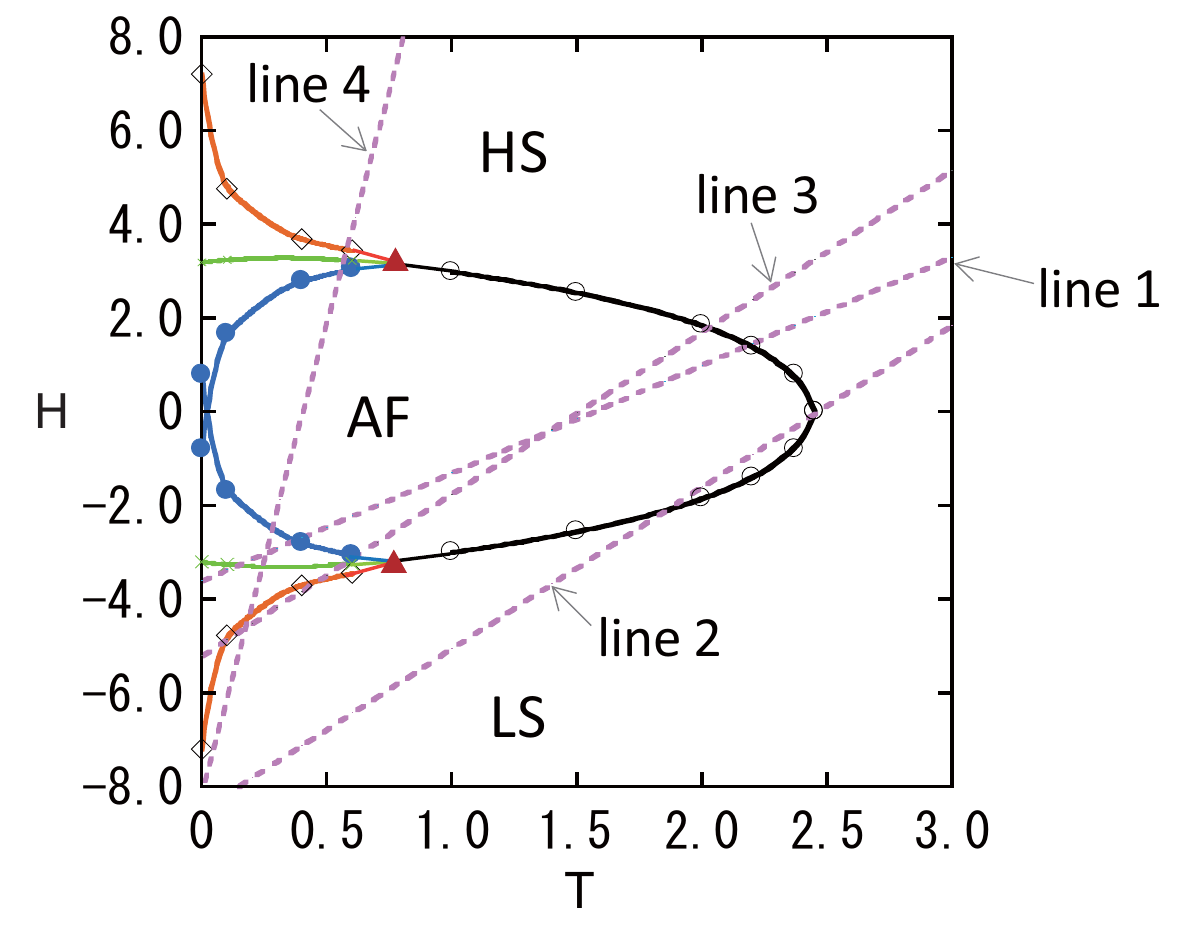}} 
\caption{ 
(color online) 
Phase diagram for the weak interaction case of the elastic model. 
The black line denotes the critical line. Orange and blue lines are spinodal lines for AF-like  and F-like phases, respectively. Green lines denote the coexistence lines of the AF-like and F-like (HS or LS) phases. Red triangles are tricritical points. Lines 1$-$4 (four dashed lines) are paths for typical SC transition patterns (see Fig.~\ref{Fig2_fhs_weak_int}).}
\label{Fig1_phase_diag_weak_elastic}
\end{figure}

\section{Phase diagram}
\label{sec_phase_diag}

In the present model, the magnetization ($m$) and staggered magnetization ($m_{\rm sg}$) are the essential order parameters. 
We define the ``magnetization" ($m$) and ``staggered magnetization" ($m_{\rm sg}$) to show F-like and AF-like order, respectively (not real magnetic order): 
\begin{equation}
m=\frac{1}{N} \Big \langle \sum_i \sigma_i \Big \rangle \;\;\; {\rm and} \;\;\; m_{\rm sg}=\langle | \tilde{m}_{\rm sg} | \rangle =\frac{1}{N} \Big \langle | \sum_i (-1)^{i_x+i_y} \sigma_i | \Big \rangle, 
\end{equation}
respectively. Here ($i_x$, $i_y$) is the integer coordinate of the $i$th molecule. The HS fraction is given by 
\begin{equation}
f_{\rm HS}=\frac{m+1}{2}. 
\end{equation}

\subsection{Relatively weak elastic interactions}
\label{sec_phase_diag_weak}

We show the phase diagram in Fig.~\ref{Fig1_phase_diag_weak_elastic} for 
the case of relatively weak elastic interactions ($k_1=400$, $k_2=40$). 
The black line denotes the critical line between the AF-like and disordered (D) phases. 
For identification of the critical points, we plotted the crossing points of the fourth-order Binder cumulants~\cite{Binder} for the AF-like order parameter for different system sizes, 
\begin{equation}
U_4^{\rm AF}(L)\equiv 1-{\langle \tilde{m}_{\rm sg}^4\rangle_L\over 3\langle \tilde{m}_{\rm sg}^2 \rangle^2_L}. 
\label{Binder cum}
\end{equation} 

In the previous study~\cite{Nishino3} at $H=0$, the critical points 
correspond to the Ising fixed-point value of the cumulant $U*=0.61\cdots$~\cite{IsingCum}.  It should be noted that the effective magnetic field ($H$) is not the conjugate field for the staggered magnetization, and the critical line exists in a range of $H \neq 0$. 
Here the value of the crossings also corresponds to the Ising fixed-point value.  
The critical temperature at $H=0$ is $T_{\rm c}=2.45$, slightly higher than that of the pure AF Ising model~\cite{Onsager}: $T_{\rm c}=\frac{2}{\ln(1+\sqrt{2})}\simeq 2.27$. This indicates that the elastic interaction slightly enhances the AF-like short-range interaction~\cite{Nishino3}. 


The orange and blue lines are spinodal lines for AF-like and F-like phases, respectively, which were estimated in a system with $L=40$. 
The green lines denote the coexistence lines of the AF-like and F-like (HS or LS) phases.
Here, we locate the coexistence lines by the following mixed start method. 

First, we prepare an initial configuration consisting of two slabs, one in the AF-like state and one in the F-like state. 
We set the F-like spin configuration, i.e., $\sigma_i=1$ (or $-1$) at the coordinate ($i_x$, $i_y$) for $1 \le i_x \le L/2$ and $1 \le i_y \le L$ and the AF-like spin configuration, i.e., alternating 1 and $-1$ for $\sigma_i$ for $L/2 \le i_x \le L$ and $1 \le i_y \le L$. 
The volumes of the AF-like and F-like phases are different, and to realize a compromise lattice state, we equilibrated the lattice for the fixed spin configuration in the two slabs at the corresponding temperature $T$. 
Then we searched for the field ($H$) at which the final state would be either with approximately 50\% probability. 
We observed that the transition region of the field $H$ between the final F-like state with 100\% probability and the final AF-like state with 100\% probability is very narrow, and the error bar of the coexistence lines is about the width of the lines. 
In this mixed start method, the final state was the F-like or AF-like state, and we did not find that the system is trapped in any other state. We adopted this method as a convenient method to estimate coexistence lines, but this method does not always give the true coexistence lines. We will discuss this point further in the Summary section.

\begin{figure}
\centerline{\includegraphics[clip,width=12.0cm]{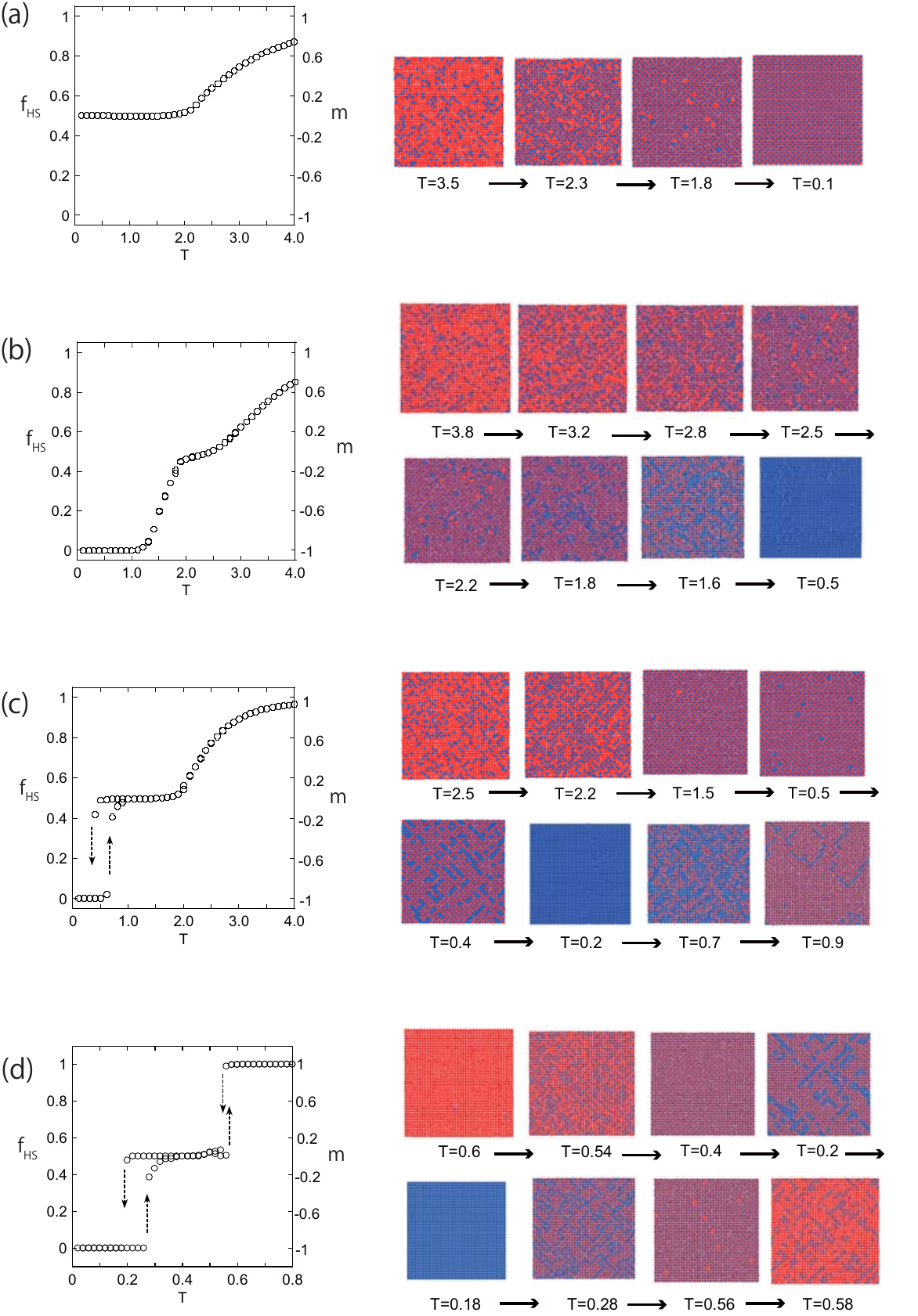}} 
\caption{ 
(color online) Temperature dependence of $f_{\rm HS}$ ($m$) and 
snapshots of the configuration for the path of (a) line 1, (b) line 2, (c) line 3, and (d) line 4 in Fig.~\ref{Fig1_phase_diag_weak_elastic}. 
The temperature changes from $T=4$ to $T=0.2$ to $T=4$ for (a)$-$(c) and 
from $T=0.8$ to $T=0.02$ to $T=0.8$ for (d).  }
\label{Fig2_fhs_weak_int}
\end{figure}

By using the phase diagram, we can easily understand the different patterns of SC transitions. 
We investigate four typical patterns~\cite{Gutlich_book,SC_book2,Gutlich}: (I) one-step SC between AF-like and HS phases, (II) two-step SC with double second-order (continuous) transitions, (III) two-step SC with first-order and second-order transitions, and (IV) two-step SC with double first-order (discontinuous) transitions. 
The parameters $D$ and $g$ are set as (case I) $D=7.2$, $g=100$, (case II) $D=17$, $g=1000$, (case III)  $D=10.4$, $g=1000$, and (case IV) $D=16$, $\ln g=40$. 
With the temperature variation, the state of the system changes along 
the paths of lines 1, 2, 3, and 4 in Fig.~\ref{Fig1_phase_diag_weak_elastic} for cases I, II, III, and IV, respectively. 

We give in Fig.~\ref{Fig2_fhs_weak_int} the temperature dependences of $f_{\rm HS}$ ($m$) and snapshots of the configurations for cases I, II, III, and IV in Figs.~\ref{Fig2_fhs_weak_int} (a), (b), (c), and (d), respectively. 
The initial state is the HS phase at $T=4$ for (a)$-$(c) ($T=0.8$ for (d)) and the temperature is lowered to $T=0.2$ for (a)$-$(c) ($T=0.02$ for (d)) and then raised back to $T=4$ for (a)$-$(c) ($T=0.8$ for (d)).

In case I, in lowering $T$ the HS phase changes to the AF-like phase at around $T=2.2$, at which line 1 in the phase diagram crosses the critical line. 
For lower temperatures, the AF-like phase remains down to $T=0$, at which the system is trapped in the metastable AF-like phase. In the successive heating process the HS fraction overlaps with that of the cooling process. Case I corresponds to one-step continuous SC transition in the Ising universality class between the HS and the AF-like phases. 

In case II, there are two crossing points between line 2 and the critical line in the phase diagram, and double continuous SC transitions occur. 
The first transition is located at $T \simeq 2.5$ between the HS phase and the AF-like phase, and the second one takes place at $T \simeq 1.8$ between the AF-like phase and the LS phase. In the transitions, clustering of AF-like phase occurs at the higher $T_c$ and F-like (LS) clustering occurs at the lower $T_c$ because of the Ising universality.  
The HS fractions during heating and cooling overlap. 

In case III, the first crossing is located at around $T=2.0$ (continuous transition) but line 3 crosses the AF-like spinodal line at $T \simeq 0.4$, and a discontinuous (first-order) transition between the AF-like phase and the LS phase takes place. In the warming process, the LS phase changes discontinuously to the AF-like phase at around $T=0.7$, at which line 3 crosses the F-like spinodal line (blue line). Here a hysteresis loop of the HS fraction is realized. It is typical for discontinuous and continuous two-step SC transitions. 
In the relaxation from the AF-like phase to the F-like (LS) phase, clustering 
of the LS phase does not take place, and this is also observed from the LS to 
the AF-like phase. This behavior indicates that a long-range interaction causes these transitions. We examine this point further in the next section. 

In case IV, the path of line 4 crosses four spinodal lines at lower temperatures than the two tricritical points. Here double first-order transitions occur and double hysteresis loops are observed.   
In lowering $T$, the HS phase changes to the AF-like phase at $T \simeq 0.54$ and then to the LS phase at $T \simeq 0.2$, which corresponds to crossing the upper F-like spinodal line and the lower AF-like spinodal line, respectively.   
In warming, the LS phase changes to the AF-like phase at $T \simeq 0.28$ and then to the HS phase at $T \simeq 0.58$, corresponding to crossing the lower F-like spinodal line and the upper AF-like spinodal line, respectively. It is worth noting that in all the processes, no clustering is observed. 

\begin{figure}
\centerline{\includegraphics[clip,width=10.0cm]{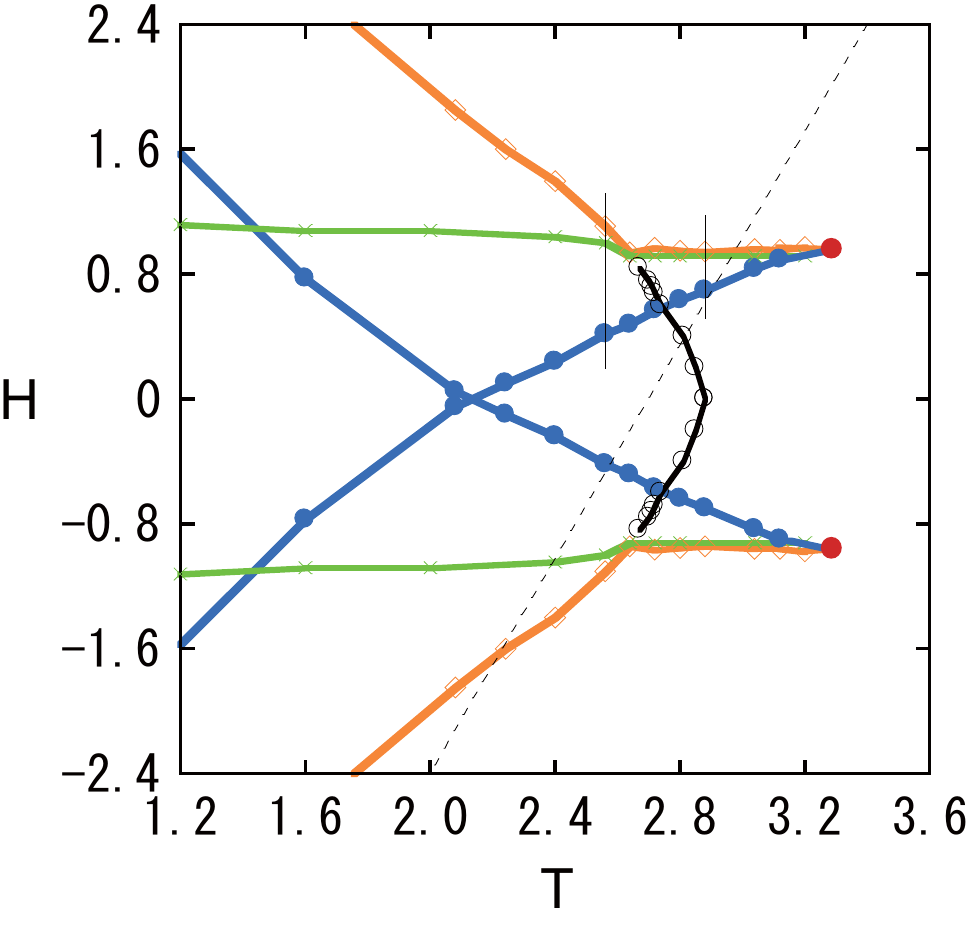}}
\caption{ (color online) Phase diagram in the relatively strong interaction case. Orange lines denote AF-like spinodal lines and D spinodal lines. 
 Blue lines are F-like spinodal lines. 
Black and green lines are the critical line and the coexistence lines, respectively. Red circles are mean-field critical points. 
Thin lines are guides for $T=2.56$ and $T=2.88$. The dashed line is a hysteresis path.}
\label{Fig_phase_diag_strong}
\end{figure}

\subsection{Relatively strong elastic interactions}
\label{sec_phase_diag_strong}

Next, we investigate the phase diagram for the relatively strong elastic-interaction case  ($k_1=1600$).   
The qualitative features in the relatively low-temperature region are essentially the same as in the relatively weak elastic-interaction case. However, in the relatively high temperature region, qualitatively different features are found. 
Thus, we focus on the latter.

The phase diagram at higher temperatures is shown in Fig.~\ref{Fig_phase_diag_strong}. The critical temperature at $H=0$ is $T_{\rm c}=2.88$, slightly higher than that of the case for relatively weak elastic interactions. 
Horn regions appear at higher temperatures, which is similar 
to the Ising antiferromagnet with moderate HT interactions. 
\begin{figure}
\centerline{\includegraphics[clip,width=11.5cm]{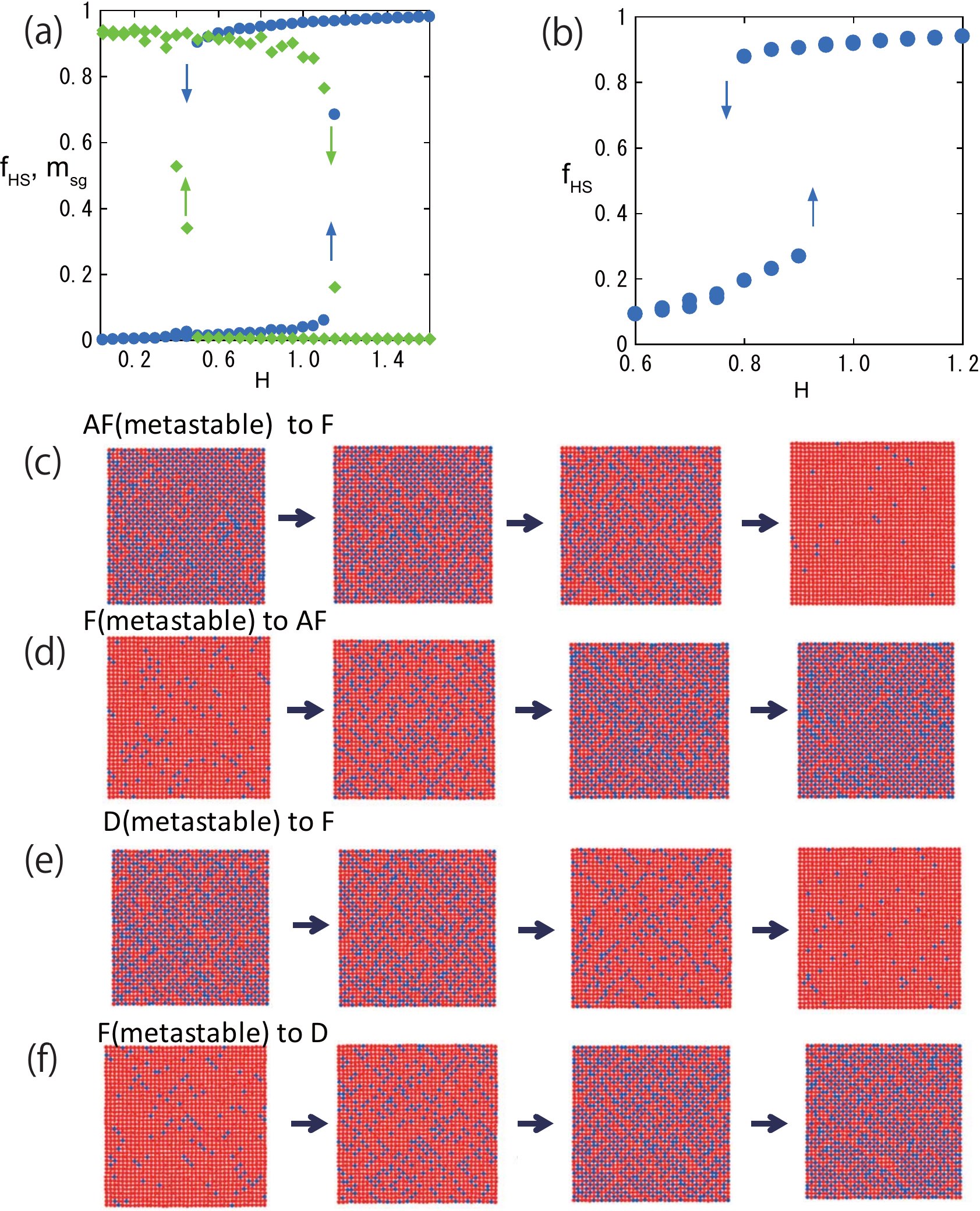}}
\caption{ (color online) (a) Field ($H$) dependence of $m_{\rm sg}$ (diamonds) and $f_{\rm HS}$ ($m$) (circles) at $T=2.56$.  (b) Field ($H$) dependence of $f_{\rm HS}$ ($m$) at $T=2.88$. The value of $H$ changes upward then downward along a thin line at $T=2.56$ for (a) and $T=2.88$ for (b) on the phase diagram (Fig.~\ref{Fig_phase_diag_strong}). (c) Snapshots from the AF-like metastable phase to the F-like phase for (a). (d) Snapshots from the F-like metastable phase to the AF-like phase for (a). (e) Snapshots from the D metastable phase to the F-like phase for (b). (f) Snapshots from the F-like metastable phase to the D phase for (b). }
\label{Fig_hysteresis_loop}
\end{figure}
Hysteresis curves of $f_{\rm HS}$ ($m$) and $m_{\rm sg}$ at $T=2.56$ 
are plotted in Fig.~\ref{Fig_hysteresis_loop} (a), and snapshots of the configurations around the AF-like spinodal point and the F-like spinodal point are shown in Figs.~\ref{Fig_hysteresis_loop} (c) and (d), respectively. 
We do not find any cluster formation of AF-like or F-like domains. Rather, the system remains uniform throughout the phase changes. This suggests that the transitions are induced by long-range interactions. 
This character is also found in the transition at the D-spinodal point at 
$T=2.88$. 
A hysteresis curve of $f_{\rm HS}$ ($m$) at $T=2.88$ is shown in Fig.~\ref{Fig_hysteresis_loop} (b), and snapshots of the configuration around the D-spinodal point and the F-like spinodal point are given in Figs.~\ref{Fig_hysteresis_loop} (e) and (f), respectively.  

\begin{figure}
\centerline{\includegraphics[clip,width=15.0cm]{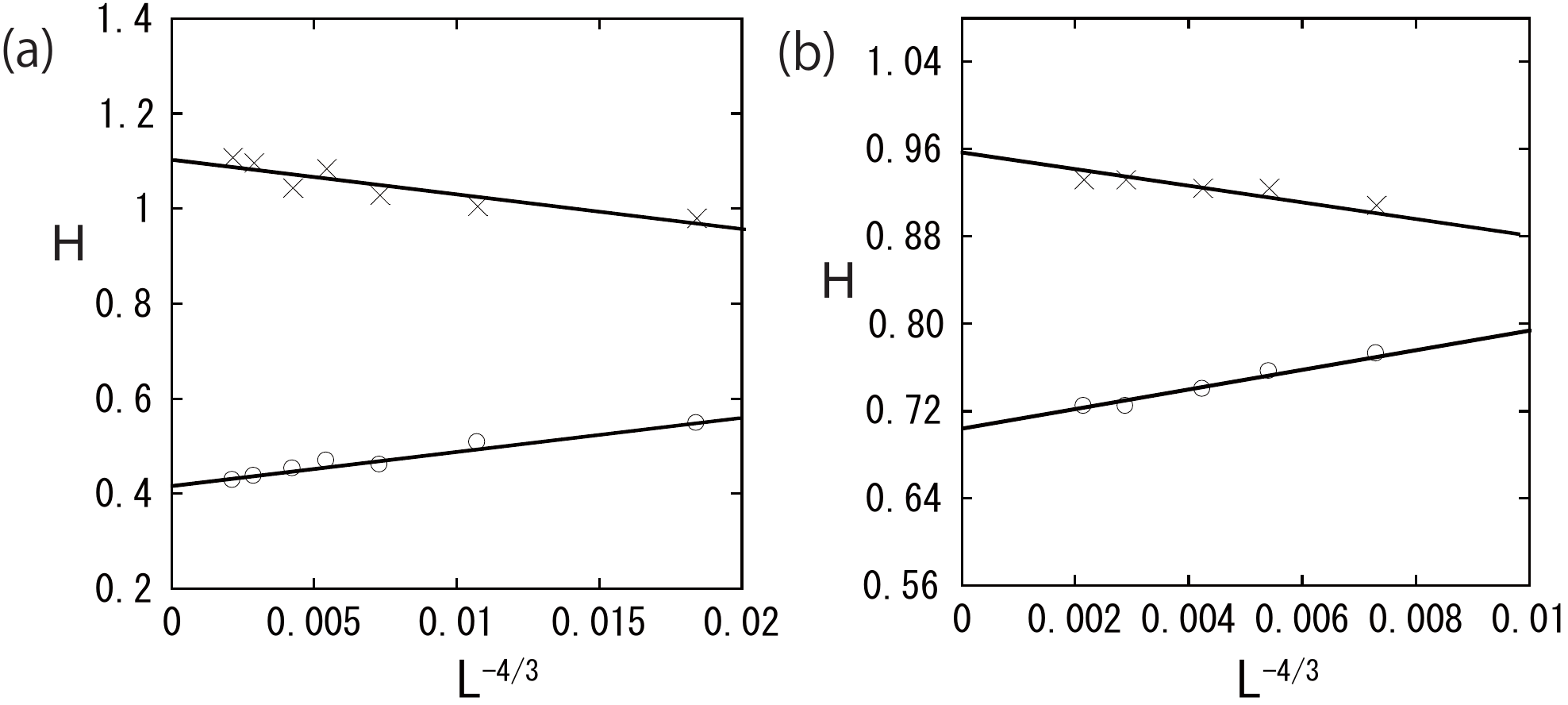}}
\caption{ (color online) System-size ($L$) dependence of the field $H_L$ for the limit of the metastable phase. (a) Crosses denote $H_L$ for the metastable AF-like phase and circles the metastable F-phase. $T=2.56$. (b) Crosses denote $H_L$ for the metastable D phase and circles the metastable F-like phase. $T=2.88$.}
\label{Fig_scaling}
\end{figure}

In short-range interaction models, nucleation and growth of droplets of the
stable phase are characteristic of the decay of the metastability. 
The metastable lifetime becomes shorter for larger system sizes and  
finally system-size independent in the thermodynamic limit~\cite{Per2}.
In contrast, the long-range, ferromagnetic interactions cause 
an exponential size divergence of the lifetime~\cite{Mori,Tomita}
 and a true metastable phase is realized. Simultaneous and discontinuous jumps occur in scanning $H$ across the limit of metastability. 
In that case, the following finite-size scaling relation holds~\cite{Paul,Miya2}. 
For the field $H_L$ marking the limit of the metastable phase at system size $L$, 
\beq
| H_{\rm spinodal} - H_L | \propto L^{-4/3}, 
\eeq
where $H_{\rm spinodal}$ is the spinodal field in the thermodynamic limit. 

For the Ising antiferromagnet with HT long-range interactions, 
this relation is well satisfied for all the spinodal lines.  
Here we analyze the scaling relation for the elastic interaction model. 

The system-size dependence of $H_L$ is plotted in Fig.~\ref{Fig_scaling} at (a) $T=2.56$ and (b) $T=2.88$. 
In Fig.~\ref{Fig_scaling} (a), the crosses denote $H_L$ for the AF-like phase and the circles  $H_L$ for the F-like phase, and in Fig.~\ref{Fig_scaling} (b), the crosses denote $H_L$ for the D phase and the circles  $H_L$ for the F-like phase. 
We find that the scaling relation is satisfied for the spinodal fields. 
In Fig.~\ref{Fig_phase_diag_strong} the AF-like (orange), F-like (blue), and D (orange) spinodal lines are given as $H_{\rm spinodal}$ obtained from this scaling relation. 

This analysis shows that all transitions between metastable and stable states are caused by the long-range interactions of elastic origin. 
On the other hand, the elastic interactions are negligible, and 
the short-range interaction ($J$) is essential for the critical line.  
This is summarized in Table I.

Next we consider the mechanisms of the different clustering features and 
the generation of the horn structures. 
Near the critical line, the average fractions of HS (LS) molecules in the AF-like phase and in the D phase are almost equal, and the density of molecules changes continuously across the transition. 
Furthermore, the interface free energy between the two AF-like phases, 
i.e., HS, LS, HS, LS, $\cdots$ and LS, HS, LS, HS, $\cdots$ and that between 
the AF-like phases and the D phase are of $O(L)$~\cite{Nishino3}. 
Exactly at the critical line, the surface tension vanishes. 
The universality class is the Ising one, and critical clustering occurs as usual in models with short-range interactions.

On the other hand, in the transitions between the F-like and AF-like phases,  
the densities of the molecules change. 
The corresponding volume change is of $O(N)$ ($N=L^2$), and during the transition the interface energy is of $O(N)$~\cite{Nishino3, Macro_nuc} as a result of the long-range interaction of elastic origin. 
Therefore, clustering is suppressed, and a mean-field-like uniform configuration change is observed. 
The same mechanism holds for the transitions between HS and LS phases (the F-like phases).

The F-like and D phases become indistinguishable at temperatures above the mean-field critical points. This situation is similar to the liquid and gas phases in the gas-liquid phase transition. 
The density of the molecules (HS fraction) changes discontinuously at the F-like and D spinodal points on the low-temperature side of the mean-field critical points. 
The D phase has no long-range order but shows strong AF-like short-range correlations. The long-range effect of the elastic interaction is essential for these discontinuous changes and mean-field-like, spatially uniform phase changes. 

For weaker elastic interactions, the phase diagram is characterized by tricritical points.  
The long-range interaction of elastic origin causes the F-like and AF-like spinodal lines, which are mean-field like and very resistant against thermal fluctuations.
The locations of the tricritical points shift to higher temperatures as 
the elastic interaction becomes stronger, while the locations of the critical points $T_{\rm c}$ at around $H=0$, which are caused by the short-range interaction, change relatively little. 

If the elastic interaction is even stronger, it causes  
the metastability of the D phase and extensions of the AF spinodal lines which accompany extensions of the F spinodal lines.   
Here the tricritical points cannot exist any longer and change to mean-field critical points. 
Horn structures accompany this qualitative change. 

In the phase diagrams for the Ising antiferromagnet with HT long-range interactions~\cite{Per1}, the tricritical points decompose into pairs of critical end points and mean-field critical points. 
Although in the present study the coexistence lines and the D spinodal lines are located very close together and the critical end points are hard to indentify precisely, the same scenario is likely to hold.

\begin{figure}
\centerline{\includegraphics[clip,width=7.0cm]{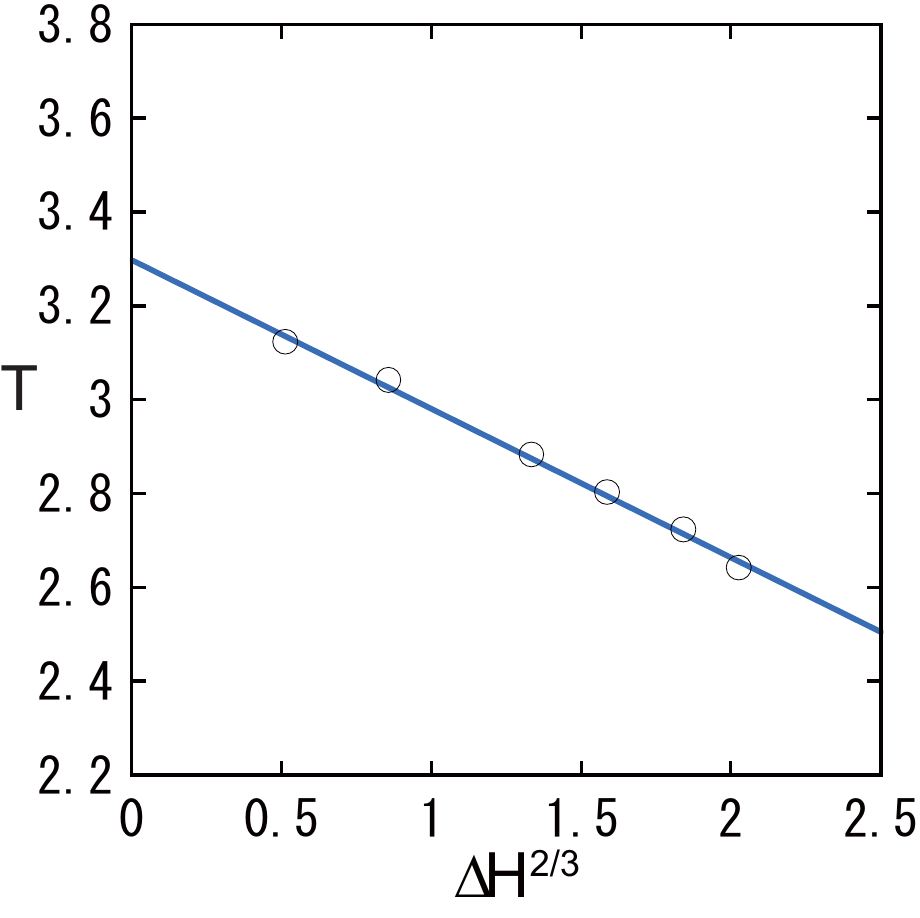}}
\caption{ (color online) Temperature vs. $\Delta H^{2/3}$ in the horn region.}
\label{Fig_scaling_dh}
\end{figure}

We may determine the temperature for the mean-field critical point 
by employing the scaling relation for the spinodal and coexistence fields 
for long-range interaction models~\cite{Per1,Newman}: 
\beq
|H_{\rm spinodal} - H_{\rm Coex}| \simeq (T_{\rm 0}-T)^{3/2}, 
\eeq
which holds for the spinodal fields near the critical, tricritical, or spinodal temperature. 
This equation leads to 
\beq
\Delta H \equiv H^{+}_{\rm spinodal} - H^{-}_{\rm spinodal} \simeq (T_{\rm 0}-T)^{3/2}, 
\eeq
where $H^{+}_{\rm spinodal}$ and $H^{-}_{\rm spinodal}$ are spinodal lines facing each other across the coexistence line, and $T_{0}$ is a tricritical point or critical point. 
Applying the D and F-like spinodal lines for $H^{+}_{\rm spinodal}$ and $H^{-}_{\rm spinodal}$, respectively, temperature vs. $\Delta H^{2/3}$ is depicted in Fig.~\ref{Fig_scaling_dh}. We find that the scaling relation holds and the mean-field critical point is located at $T_{0} \simeq 3.3$.

\begin{figure}
\centerline{\includegraphics[clip,width=8.0cm]{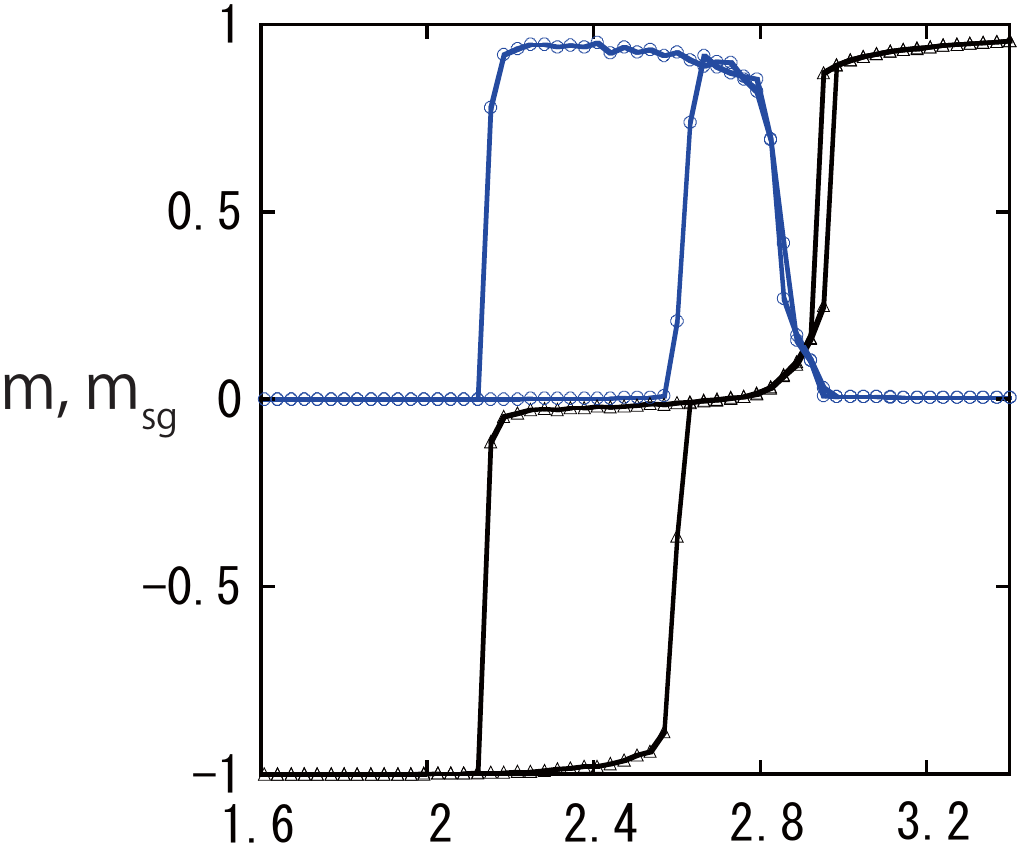}}
\caption{ (color online) Temperature dependence of $m$ (black line) and $m_{\rm sg}$ (blue line) along the hysteresis path in Fig.~\ref{Fig_phase_diag_strong}. }
\label{Fig_hysteresis_path}
\end{figure}

Here we show a new pattern of SC transitions, which is different from patterns (I)-(IV). 
Because of the appearance of the horn regions, we see a change of the phase: HS $\rightarrow$ D $\rightarrow$ AF $\rightarrow$ LS $\rightarrow$ AF $\rightarrow$ D $\rightarrow$HS along 
the hysteresis path with $D=18.514$ and $g=950.647$ in the phase diagram in  Fig.~\ref{Fig_phase_diag_strong}. 
We plot the temperature dependence of $m$ and $m_{\rm sg}$ for $L=50$ along the hysteresis path (Fig.~\ref{Fig_hysteresis_path}). The temperature is lowered and then raised. 
Initially the HS state is realized and it changes discontinuously to the D state at around $T \simeq  2.9$. At this temperature, the AF-like order already develops in the D state due to finite-size effects. 
Then, the D state changes continuously to the AF-like state, which remains until $T\simeq 2.2$, and it discontinuously changes to the LS state. 
In the returning process, the LS state changes discontinuously to the AF-like state at around $T \simeq 2.6$. Then, it changes continuously to the D phase, and discontinuously to the HS state at around $T=3.0$.

\begin{table}
\caption{Role of the elastic interaction for phase transitions}
\begin{tabular}{lcc} \hline
Transitions & Clustering  & $\;\;$ Elasticity (LR interaction)  \\ \hline
critical line: AF-like phase $\leftrightarrow$ D phase & yes & inessential (irrelevant)
  \\
spinodal line: AF-like phase $\rightarrow$  F-like phase & no  & essential  \\
spinodal line: F-like phase $\rightarrow$  AF-like phase & no & essential  \\
spinodal line: D phase $\rightarrow$  F-like phase & no  & essential  \\
spinodal line: F-like phase $\rightarrow$  D phase & no  & essential  \\
spinodal line: F-like phase $\rightarrow$ F-like phase & no & essential  \\ \hline
\end{tabular}
\end{table}

\section{Summary}
\label{summary}

We have investigated the phase diagrams and properties of phase transitions for an elastic interaction model for spin crossover materials with AF-like short-range interactions. 
For stronger elastic interactions, unusual ``horn structures," which are surrounded by the mean-field critical points, the critical line, F-like, and D spinodal lines, are realized. 
These structures are resistant to fluctuations because the spinodal lines are derived from the long-range interaction of elastic origin. 
For weaker elastic interactions,  tricritical points are realized instead of the mean-field critical points. 
The critical lines are caused by the short-range interaction and are slightly shifted due to a small enhancement of the short-range interaction by the elastic interaction. However, the nature of the criticality is not affected, and the long-range interaction of elastic origin is irrelevant (inessential).    

We have observed different features of domain growth of the new phase at the critical line and the spinodal lines. Clustering is observed around the transition temperature in the former, while no clustering occurs in the latter. 
This confirms that the short-range and long-range interactions are essential for the former and the latter, respectively. 

Similar horn structures of the phase diagram are found in the Ising AF magnet with HT long-range interactions~\cite{Per1}, and we suggest that this kind of structure is universal for systems with an interplay between competing short-range and long-range interactions. 

Here we adopted the mixed start method to identify the coexistence lines in the horn structures. In very recent work for the Ising antiferromagnet with HT long-range interactions~\cite{Per3,Per4}, a Wang-Landau MC sampling method was applied to investigate the detailed structure of the phase diagram, and a modification of the locations of the coexistence lines in the horn region was pointed out. They were found to be located in the middle between the spinodal lines, in contrast with the results obtained by the mixed start method. 
The reason for the difference may be attributed to a dynamical effect, but remains a problem for future study. 
The coexistence line could be located in the same way in the case of the elastic SC model. 
The accurate identification of the coexistence line in this model will be considered in the future.

\section*{Acknowledgments}
The present work was supported by Grants-in-Aid for Scientific Research C (17K05508, 26400324 and 25400391) from MEXT of Japan. 
The authors thank the Supercomputer Center, the Institute for Solid State Physics, the University of Tokyo, for the use of the facilities. 
P.A.R. gratefully acknowledges hospitality by the Department of Physics, University of Tokyo, and partial support by U.S. National Science Foundation Grant No. DMR-1104829.


\end{document}